\documentclass[amssymb,amsmath,aip,jcp,preprint]{revtex4-1}
\usepackage{amsfonts,eucal,bm}
\usepackage{graphicx}
\usepackage{color}
\usepackage{rotating}
\usepackage{subfigure}
\usepackage{hyperref}
\usepackage{cleveref}

\def\sss#1{  {\scriptscriptstyle\rm #1} } 

\mathchardef\hyph="2D
\DeclareMathAlphabet{\matheurm}{U}{eux10}{m}{n}
\def \bea { \begin{align} }

\def \be { \begin{equation} }

\def \ee { \end{equation} }

\def \eea { \end{align} }

\begin{document}

\author{Sergey N. Maximoff}
\email[To whom correspondence should be addressed: ]{SNMaximoff@gmail.com}
\affiliation{Department of Chemistry, University of California, Berkeley,  USA}
\affiliation{Department of Chemistry and Biochemistry, Loyola University, Chicago, IL, USA}
\title{Role of charge transfer in catalytic hydrogen oxidation over platinum}

\begin{abstract}

Charge transfer plays the central role in  chemisorption
 and chemical reactions at metal surfaces. 
The Somorjai group in Nano. Lett. {\bf 9}, 3930 (2009) has reported a chemicurrent,  a flux of  charge carriers 
in response to $\mathrm{H_2}$ oxidation by $\mathrm{O_2}$ into $\mathrm{H_2O}$ over a platinum surface. 
The  nature of the chemicurrent has been debated
in the literature; explanations both extrinsic and intrinsic to the reaction mechanism have been offered.
This article suggests a picture behind the chemicurrent experiments, which is	
 intrinsic  to the mechanism of hydrogen oxidation in a high   temperature regime.  Surface reaction intermediates, $``\mathrm{O^-}"$, $``\mathrm{H^-}"$, $``\mathrm{OH^-}"$, are negatively charged while  the product, $\mathrm{H_2O}$, and reactants,   $\mathrm{H_2}$ and $\mathrm{O_2}$,  are neutral. Hence, charge transfers between the metal and reacting species are inevitable. After electrons are transferred from the metal to the interface in dissociative oxygen and hydrogen adsorption, translationally excited electrons in the metal arise in part as a result of decay  of negatively charged transition states  in reaction steps connecting surface species of different charges,  $``\mathrm{O^-+H^-}" \rightleftarrows ``\mathrm{2 e+OH^-}"$ and $``\mathrm{OH^-+H^-}"\rightleftarrows ``\mathrm{2 e+H_2O}"$. These transition states   are the lowest energy configurations possible for changing the charge of the negatively charged surface intermediates. In particular, the transition state for $``\mathrm{O^-+H^-}" \rightleftarrows \mathrm{``2 e+OH^-"}$ limits the reaction rate. 
 These processes may contribute to the chemicurrent.
This picture should also apply to other catalytic chemical reactions on metal surfaces that
proceed through formation of charged surface intermediates and their consequent discharge into the metal.  
\end{abstract}
\maketitle

\section{Introduction}

\begin{figure}[h]
\subfigure[\label{Fig:InterfaceA}]{
\includegraphics[scale=0.45]{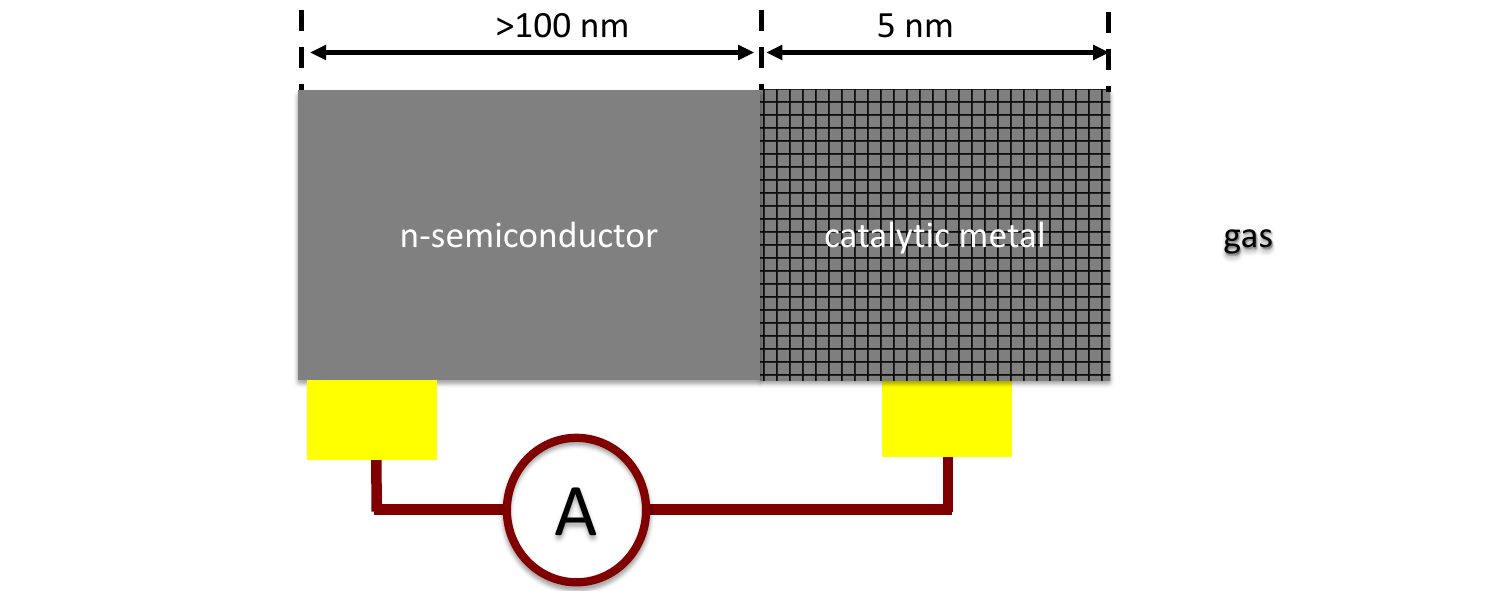}
}\\
\subfigure[\label{Fig:InterfaceB}]{
\includegraphics[scale=0.45]{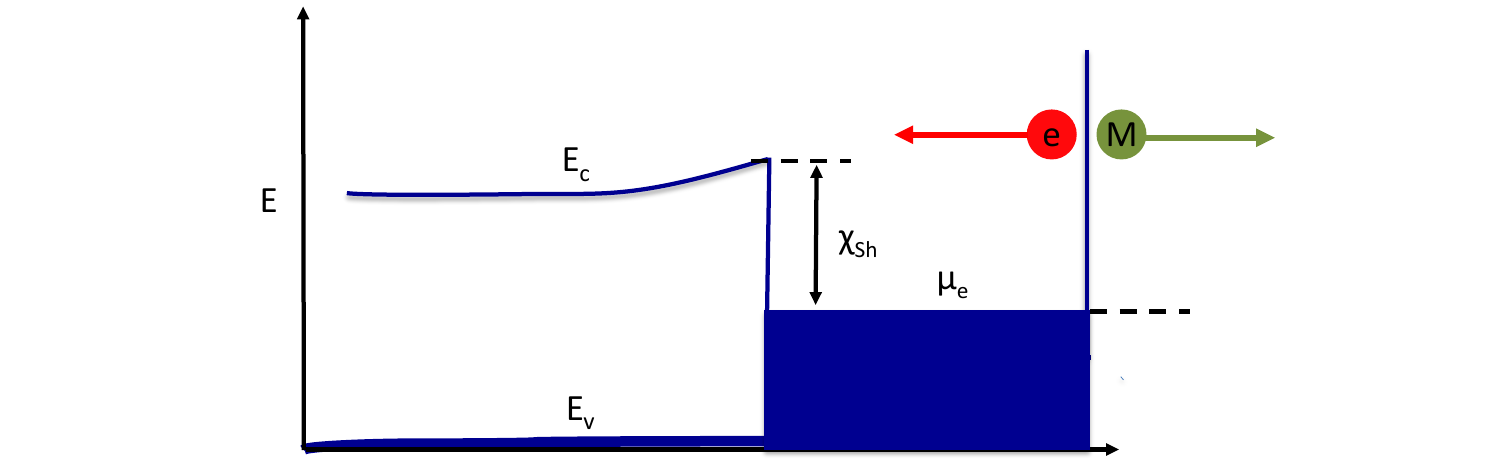}
}
\caption
{
(a) The Schottky diode. 
The chemistry takes place at the interface 
 of the gas and the catalytic metal (right),
   the catalytic metal joins  the $n$-semiconductor over the Schottky 
contact.
The Ohmic contacts (the yellow bars) collect the charge from the catalytic metal and
the semiconductor. The wires (brown lines) close the circuit 
through the ammeter. 
(b) An energy diagram of the Schottky diode. 
The chemical potential of the electrons ($\mu_{\sss{e}}$)
 is below the Schottky barrier ($\chi_{\sss{Sh}}$).
If a chemical process excites a charge whose kinetic energy 
at the Schottky contact exceeds the Schottky barrier, the charge enters
the semiconductor where it travels in the conduction band ($E_{\sss{c}}$) above the valence band ($E_{\sss{v}}$). 
The workfunction  and the absorption barriers  (the rightmost blue vertical line) prevent 
the electrons from spilling into the gas and the molecular species 
from entering the metal, respectively. 
 }
\end{figure}

Absorption of light  in metals may produce hot internal electrons  whose kinetic energies exceed typical thermal values \cite{PETEK1997239}.  
Exothermic chemistry at low workfunction metal surface may eject electrons from
metal in the form of exoelectrons \cite{PhysRevLett.65.2035,Nienhaus:2002}.
 Evidence has been mounting
\cite{ISI:000271566400045,ISI:000298240000014,Nienhaus:2002,Auerbach:2000,
ISI:000271468000037,Nienhaus:2001,ISI:000282855400055,ISI:000315595700003,ISI:000317032100039,Nienhaus:1999,doi:10.1021/cr400311p}
 that exothermic
redox  chemical reactions at
high workfunction surfaces of electron-rich noble  and late transition  metals
  may   also invoke hot electron excitations that  propagate    
 for about a few nanometers into the metal at  kinetic energies
 $\gtrsim 0.5-1.2\;\mbox{eV}$, but below 
the workfunction barrier for electron ejection from the metal ($4.3-5.9\;\mbox{eV}$\cite{LB:III:42A2}).
These excitations must propagate inside the metal rather than in a vacuum,
 and therefore are tricky to probe  directly in  experiments.
Methods of their detection are  based on  Schottky diode  experiments (\cref{Fig:InterfaceA,Fig:InterfaceB}).
These experiments aim to register  these excitations indirectly
as a zero voltage bias electric current through  
 high workfunction metal film/semiconductor Schottky  diodes 
with $0.5-1.2\;\mbox{eV}$ Schottky barriers \cite{Nienhaus:2001,ISI:000271566400045,ISI:000256846000001} 
during adsorption-desorption  or reactive
adsorption-desorption of molecules or atoms  at the interface between the gas and the diode
 at  pressures ranging from ultra high vacuum (UHV)  to 
 pressures on the order of one bar \cite{ISI:000271566400045}.
 A catalytic Schottky diode  is a thin film of a catalytic metal that joins 
a thicker film of a semiconductor so that the Schottky barrier
 at the interface of the metal 
and semiconductor passes the current of thermal electrons
 in the direction from the semiconductor to the metal but not in the opposite direction. 
 However, a current originating  at the surface and passing through the metal film may circumvent
 the Schottky barrier and enter the semiconductor  when  the kinetic energy of carriers exceeds the Schottky barrier height. These hot carriers, which a  surface chemical reaction may generate, would then contribute to the zero bias current through the diode. 
 
The Refs.~\onlinecite{ISI:000271566400045,ANIE:ANIE201410951} have  reported
that an exothermic  catalytic  chemical process (the reaction heat of $-2.51 \;\mbox{eV}$ per an $\mathrm{H_2O}$
\cite{NIST:WebBook}) at the constant partial pressures $p_{\sss{H_2}}=6\;\mbox{Torr}$ and $p_{\sss{O_2}}=760\;\mbox{Torr}$,
\begin{align}
\mathrm{H_2+{1\over 2} O_{2}+Pt(111)}
\longrightarrow
\mathrm{H_2O +Pt(111)}, &   
\label{hydrogen_oxydation_stoi}
\end{align}
over  a $5\;\mbox{nm}$ thick $\mathrm{Pt}$ film in
$\mathrm{Pt/n\hyph TiO_2}$ diode with the Schottky barrier of $1.1-1.2\;\mbox{eV}$ induces a zero-bias current.
   Refs.~\onlinecite{ISI:000271566400045,ANIE:ANIE201410951} further separate this current into a component due to 
a temperature gradient across the Schottky diode (thermoelectric current) and another component
proportional to the reaction rate --- the chemicurrent. 
This partitioning scheme does not expose molecular mechanisms, which are responsible for the measured zero-bias current,
and it has been debated  in the literature \cite{ISI:000315595700003,doi:10.1021/jp210492k,:/content/avs/journal/jvsta/31/2/10.1116/1.4774217} whether chemicurrent
is a phenomenon that has to do with charge transfer surface chemistry at high pressure conditions. 

It is not at all  new   that internal  degrees of freedom of  molecules play a role in energy flow
between metal electrons and molecules through transient retention of metal electrons
by the internal degrees of freedom \cite{ISI:000165670900007,ISBN:9783642329548}.
Refs.~\onlinecite{Gadzuk:1983,ISI:000271468000037} bring up the significance of temporary molecular anions in energy relaxation in 
 molecular scattering against metal surfaces. 
Refs.~\onlinecite{Persson:1982,Head-Gordon:Tully:1995,Krishna:Tully:2006} discuss how transient charge transfer between 
adsorbates and metal surfaces facilitates   rovibrational energy relaxation.
A review \cite{doi:10.1142/S0218625X95000285} discusses exoelectron emission during oxygen adsorption on
alkali metal surfaces. 
It can happen in chemical reactions that the number of electrons localized at reacting intermediates changes significantly and rapidly upon crossing the activation barrier at a transition state (i.e., a saddle point at the ground state adiabatic surface) along the reaction coordinate \cite{ISI:000267972700008}. This structure of the ground state adiabatic surface conveys an instability in the number of electrons localized at the interface in the adiabatic ground state and preconditions non-adiabatic conversion of high internal energy at the transition state into kinetic energy of delocalized metal electrons. As  it turns out, this latter picture is relevant in the catalytic hydrogen oxidation reaction \cref{hydrogen_oxydation_stoi}. 

This article examines \cref{hydrogen_oxydation_stoi}  in a particular case of a high temperature regime
 when water is not present at the $\mathrm{Pt(111)}$ interface
to reveal typical events that may contribute to the chemicurrent.
In this case,
the reaction mechanism is relatively well-understood \cite{somorjai2010introduction}. In particular, it can be scrutinized theoretically
 for potential sources of chemicurrent.
First,  $\mathrm{O_2}$ and $\mathrm{H_2}$  from the gas dissociatively adsorb at the interface and
give rise to chemisorbed $\mathrm{O}$ \cite{Gland:Fisher:1980} and $\mathrm{H}$\cite{Ludwig:2008}.
Second, a surface  hydroxyl, $\mathrm{OH}$, forms.
Third, a water molecule forms upon hydrogenation of $\mathrm{OH}$.

The dissociative $\mathrm{O_2}$ adsorption on $\mathrm{Pt}(111)$
(heat of  dissociative adsorption is $-2.0\;\mbox{eV}$  per an $\mathrm{O_2}$ 
 at the saturation coverage) is a sequential electron attachment  process.
$\mathrm{O_2}$ remains physisorbed and uncharged, 
$``\mathrm{O_2}"$ (quotation marks enclose an interfacial specie), at $T\lesssim 45\; \mbox{K}$
 \cite{Luntz:1989,Wurth:1990,Puglia:1995,Artsyukhovich:Harrison:1996}, it
converts to the charged superoxide, $``\mathrm{O_2^-}"$,
 in the temperature range
$90- 135\;\mbox{K}$  \cite{Steininger:1982,Yoshinobu:Kawai:1995},
 and then to  the charged
 peroxide, $``\mathrm{O_2^{2-}}"$,
 in the temperature range
$135-150\;\mbox{K}$ \cite{Luntz:1989,Wurth:1990,Puglia:1995}, and then to the pair of 
 oxygen ions, $``\mathrm{O^-+O^-}"$,  at   higher temperatures,
 \begin{align}
\mathrm{O_2+2 e} \rightleftarrows &
\nonumber
\\
``\mathrm{O_2+2e}" 
\rightleftarrows  ``\mathrm{O_2^-+e}"\rightleftarrows & ``\mathrm{O_2^{2-}}"
\rightleftarrows ``2\mathrm{O}^-".
\label{Eq:O2Diss0}
\end{align}
The observed  increase in the electronic workfunction of $\mathrm{Pt}(111)$
 upon adsorption \cite{LB:III:42A2}, 
core level spectroscopies \cite{Wurth:1990,Puglia:1995},
 the O$-$O bond elongation, and   electronic structure calculations \cite{Eichler:1997} all
indicate that   the negative charge progressively increases from virtually none in
$``\mathrm{O_2}"$, to larger values in $``\mathrm{O_2^-}"$ and in $``\mathrm{O^-}"$, 
due to  charge transfer from the metal to  interfacial electronic states derived from
the lowest unoccupied molecular orbitals (LUMO) of $\mathrm{O_2}$.

The $\mathrm{Pt(111)}$ interface must be predominantly covered with oxygen in the excess oxygen \cite{ISI:000271566400045}.
Therefore, $\mathrm{H_2}$ dissociates on  $\mathrm{Pt(111)}$ precovered with $\mathrm{O}$
 (initial heat of adsorption is around $-0.7\;\mbox{eV}$ per an $\mathrm{H_2}$).
Dissociative $\mathrm{H_2}$ adsorption  \cite{LB:III:42A2,Ludwig:2008} on most metal surfaces results
in a charge transfer from the metal that causes formation of $``\mathrm{H}^-"$,
\begin{align}
\mathrm{H_2}+\mathrm{2 e}\rightleftarrows	
``\mathrm{H_2}+\mathrm{2 e}"\rightleftarrows 
``\mathrm{2 H}^-".
\label{Eq:H2Diss0}
\end{align}

If  $``\mathrm{H_2O}"$ is not present at the interface,
 the reaction proceeds through 
 \begin{align}
``\mathrm{O}^-+\mathrm{H}^-"\rightleftarrows
[``\mathrm{OH}^-+\mathrm{e}"]^{\sss{\neq}}\rightleftarrows
``\mathrm{OH}^-+\mathrm{e}",
\label{Eq:OHI0}
\end{align}
 at a relatively high activation energy of   $0.69\;\mbox{eV}$\cite{Anton:1990}. 
An  $``\mathrm{OH^-}"$ in \cref{Eq:H2O0}  further  recombines with a $``\mathrm{H^-}"$,
\begin{align}
``\mathrm{OH^-}+\mathrm{H}^-"\rightleftarrows&
\nonumber
\\
[``\mathrm{OH^-+H +e}"]^{\sss{\neq}}	\rightleftarrows &
``\mathrm{H_2O}+2\mathrm{e}",
\label{Eq:H2O0}
\end{align}
 at an activation energy lower than that for 
$``\mathrm{OH^-}"$ formation \cite{Volkening:Jacobi:Ertl:1999,Michaelides:2001,Karlberg:2006}.

The product, $\mathrm{H_2O}$ in \cref{Eq:H2O0},
 and the reactants, $\mathrm{H_2}$  and $\mathrm{O_2}$ in \cref{Eq:O2Diss0,Eq:H2Diss0},
 are electroneutral, but  the above discussion indicates that  \cref{hydrogen_oxydation_stoi} involves 
anionic  intermediates at the metal-gas interface.  
Since the chemical process in  \cref{hydrogen_oxydation_stoi} is catalytic,
the  anionic intermediates are not permanent, and, by charge conservation, an intermittent
charge flux  from and to the interface
is inevitable.
In what follows, this article relies on quantum chemistry models   to
show that \cref{hydrogen_oxydation_stoi} necessitates intermittent
 charge transfers from the interface to 
metal. 
 These charge transfers occur at high energy configurations  that correlate with
  the transition states $[``\mathrm{OH^-+H +e}"]^{\sss{\neq}}$ and $[``\mathrm{OH}^-+\mathrm{e}"]^{\sss{\neq}}$.
   The  barriers at these transition states  are the activation barriers for associative electron detachment  
events.  For the associative electron detachment events  to proceed,
the total energy of the colliding anionic intermediates on the left-hand sides of  \cref{Eq:H2O0}
must exceed the minimum  barrier  for  electron detachment
into the conduction band of the metal
from the interface (the middle terms in  \cref{Eq:OHI0} and \cref{Eq:H2O0}).
    
\section{Methods\label{Sect:Methods}}

The Pt(111) metal-vacuum interface
 is modeled by 4 layer hexagonal
$p(2n\times 2m)$-Pt
 slabs with $n, m=1,2$ with a single side open for adsorption.
 The spacing
 between the periodically replicated parallel 
slabs in the direction perpendicular
 to the surface is maintained above $17\; \mbox{\AA}$
 to diminish  the spurious dipole-dipole 
 interaction between the slabs that asymptotically 
decreases as the inverse cube of the separation distance.
A  long-range ordered
 honeycomb $p(2\times2)-\mathrm{O}$ pattern in register with $\mathrm{Pt}(111)$
such that every other {\it fcc} site is occupied by $``\mathrm{O^-}"$ \cite{Materer:1995} corresponds to
the $0.25$ monolayer of $``\mathrm{O^-}"$  saturation coverage upon exposure of $\mathrm{Pt}(111)$ to $\mathrm{O_2}$ at
 $T\gtrsim 150\;\mbox{K}$. This structure is taken as a model of oxygen adsorption. 

Numerical density functional theory calculations
 with periodic boundary conditions
 in this work have been done using
 the Quantum-ESPRESSO suite of computer programs  \cite{QEspresso}.
The exchange-correlation energy is given by the PW91 
generalized gradient approximation \cite{PW91}. 
 The core electrons in H, [He] O,  [Xe] Pt
 are described by ultrasoft pseudopotentials, which are fitted
 to reproduce the numerical Kohn-Sham orbitals of the atoms  \cite{QEspresso}.  
 The ultrasoft pseudopotential for Pt 
includes the scalar relativistic effects as well as the non-linear core corrections.
  The Kohn-Sham orbitals and the charge density
 of the valence electrons are expanded into the Fourier series
over  plane waves with kinetic energies below 612.26 eV and 6122.6 eV,
 respectively.

The equilibrium  
geometries are computed by
 variable cell relaxation runs
 until the total force and the stress 
falls below $0.026\;\mbox{eV/\AA}$ and $1\; \mbox{kbar}$, respectively.
Minimum energy pathways connecting the equilibrium
 structures on the adiabatic ground state surface
 are sampled by the nudged elastic bands (NEB)
 method \cite{ISI:A1995QG63800024}.
 A string of intermediate structures interpolating
 between the initial and final equilibrium structures
 is varied until forces perpendicular to the path fall below $0.094\; \mbox{eV/\AA}$.
 In the NEB and geometry optimization calculations, the integrals over
 the irreducible surface Brillouin zone of the hexagonal $p(2n\times 2m)$ 
lattice are discretized over $4/n\times 4/m$ Monkhorst-Pack grid
\cite{Monkhorst:1976} with
 the Methfessel-Paxton first-order smearing \cite{Methfessel:1989}
 of $0.204\; \mbox{eV}$.

The electronic structure at fixed configurations of atomic cores is quantified by  projected densities of states
and by Bader atomic charges \cite{Henkelman2006354}.
 The reciprocal space integrals
 in the expressions for the density of electronic states,
 the electron number density,
 and the Fermi level are computed over 
 $32/n\times 32/n$ Monkhorst-Packs grid using the tetrahedron method \cite{Blochl:1994}.
The projected density of states is computed using  projections on the valence atomic orbitals.

\section{Results and discussion\label{Sect:EmpEv}}

\begin{figure}
\includegraphics[scale=0.6]{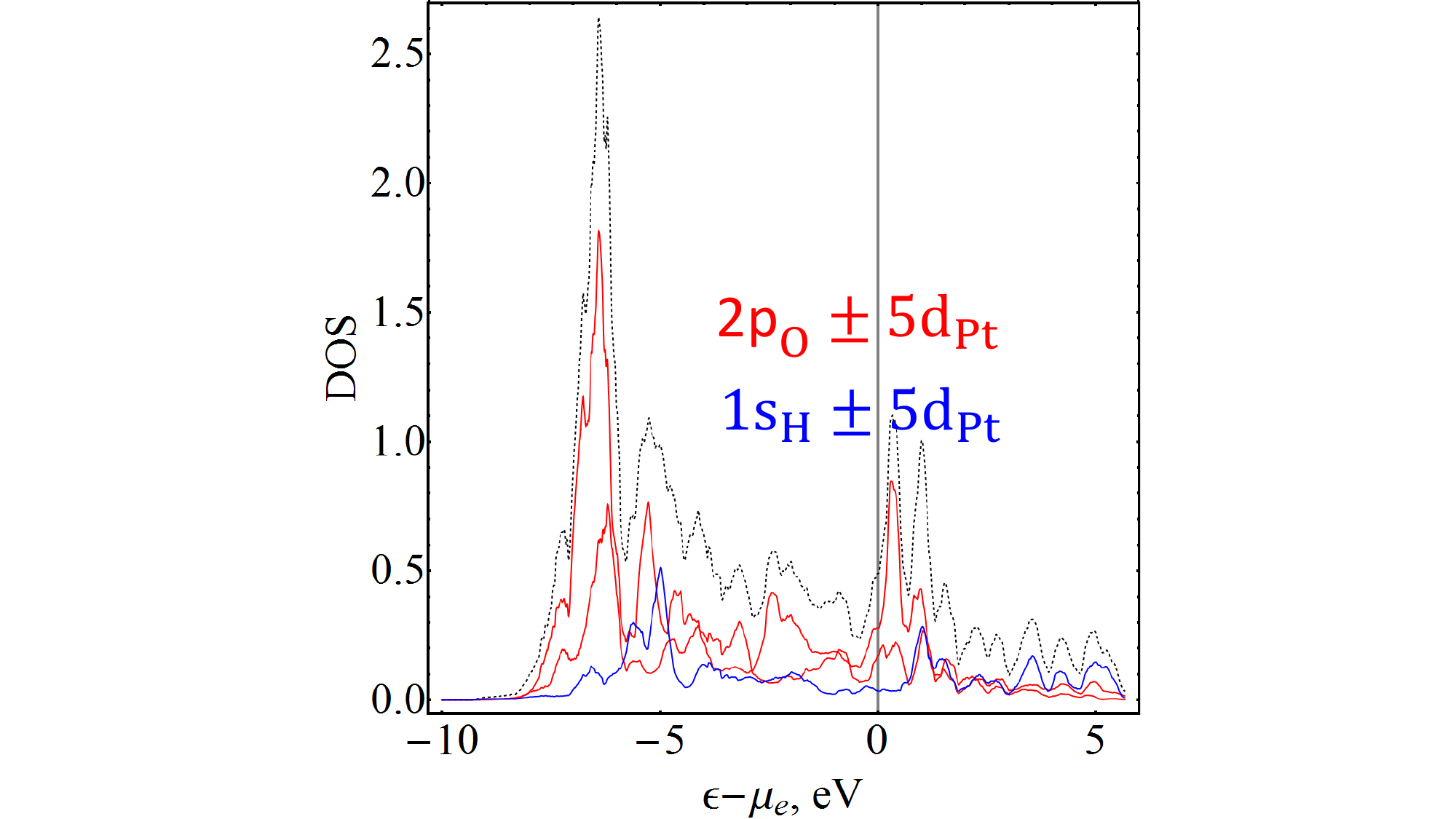}
\caption{The density of the Kohn-Sham  electronic states
	at the coadsorbed $\mathrm{O}$ and $\mathrm{H}$ ions (the black curve), 
that of the $2p_{\sss{O}\parallel}\pm 5d_{\sss{Pt}}$ states (the upper red curves),
 the $2p_{\sss{O}\perp}\pm 5d_{\sss{Pt}}$ states (the upper red curves),
that of the $1s_{\sss{H}}\pm 5d_{\sss{Pt}}$ states (the blue curve).}
\label{Fig:OandHDOS}
\end{figure}

\begin{figure}[h]
\includegraphics[scale=0.65]{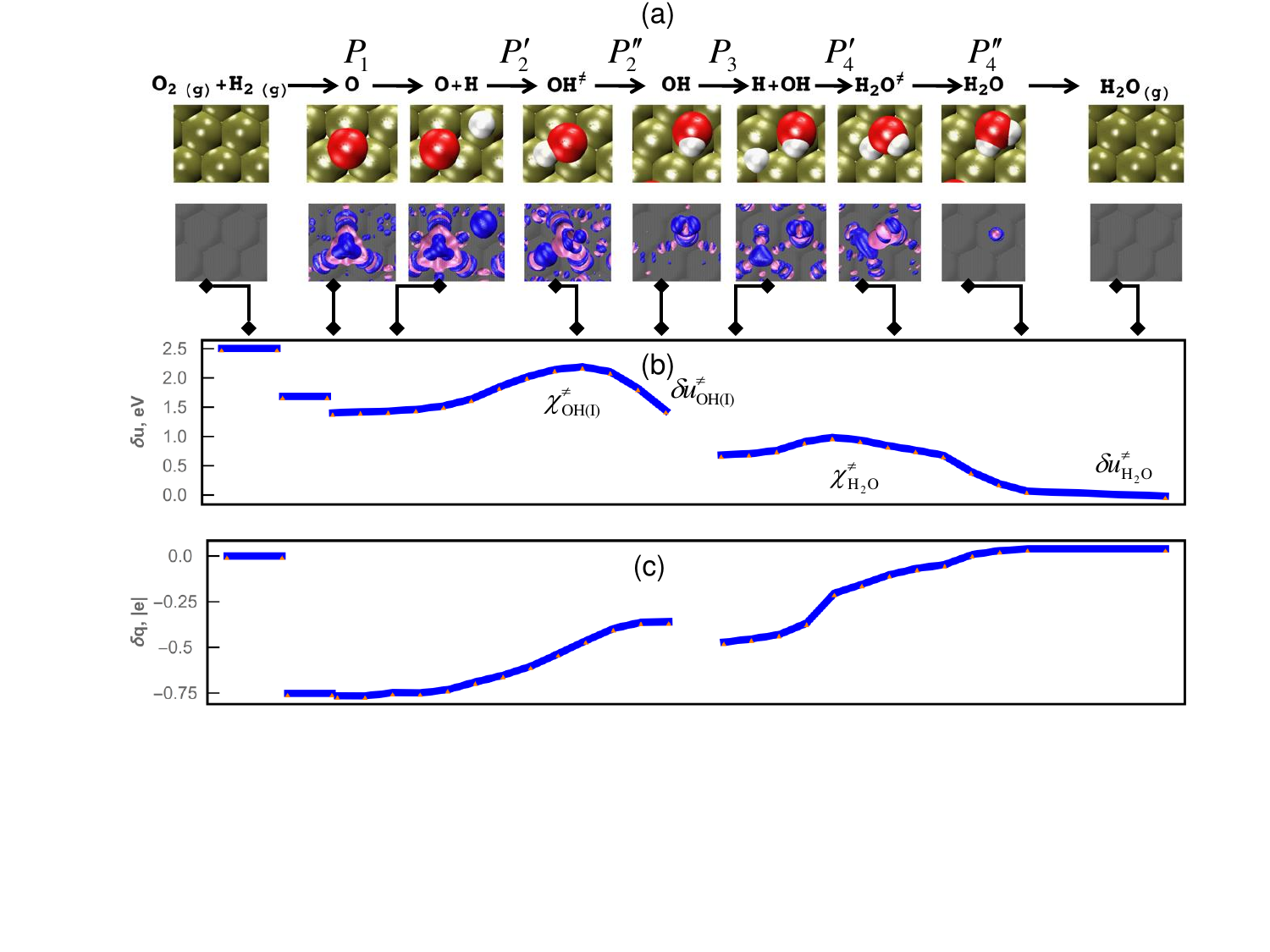}
\caption
{
Along minimum energy
 reaction paths are shown:
 (a) the surface oxygen (red balls)
 and hydrogen (white balls)
 atoms on $\mathrm{Pt}(111)$ (upper row),
 and the surface electron density $\pm0.005$
  contour values relative to that decoupled
 from the metal adsorbate layer and that of $\mathrm{Pt}(111)$  
(blue and pink represent build up and depletion of electrons, respectively); 
(b) the changes in the energy with respect to the clean $\mathrm{Pt}(111)$ 
 slab and a free $\mathrm{H_2O}$;
(c) the charge at the interface.
}
\label{Fig:ABCD}
\end{figure}

 \Cref{Fig:OandHDOS} shows the density of the electronic states at $``\mathrm{O^-}"$
in the saturation coverage
structure, $p(2\times 2)\hyph\mathrm{O_{\sss{fcc}}}$ (\cref{Fig:ABCD}(a)).
The excesses of  electrons  at the $2p_{\sss{O}}$ orbitals
and  the depletion of electrons at the $5d$ orbitals are readily seen within 
the triangle of the interfacial $\mathrm{Pt}$ sites surrounding the {\it fcc} hollow sites in
the interfacial layer, as well as in the $d_{z^2}$ orbital of the $\mathrm{Pt}$ atom
 just beneath the ${\it fcc}$ hollow site. These observations are consistent with the picture of 
$\mathrm{O_2}$ adsorption, which is established in the literature \cite{Panas1988458,LB:III:Generic}. 

 \Cref{Fig:OandHDOS} shows the density of the electronic states at $``\mathrm{H^-}"$.
 The bonding $1s_{\mathrm{H}}$-metal states are  below the Fermi level. 
The excess electron density in  \cref{Fig:ABCD}(a) indicates
 the excess of  electrons  at the $1s_{\sss{H}}$ orbitals of  $``\mathrm{H^-}"$  and the deficit of 
the electrons at the $\mathrm{Pt}$ $5d6s$ orbitals  underneath. 
The charge of an oxygen atom in  $p(2\times 2)\hyph\mathrm{O_{\sss{fcc}}}$ (\cref{Fig:ABCD}(a))  is $0.75\;\mbox{e}$.
The charges of  oxygen and hydrogen atoms in  $p(2\times 2)\hyph\mathrm{O_{\sss{fcc}}+H_{\sss{top}}}$ 
coadsorption structure
are $0.75$ and $0.02\;\mbox{e}$, respectively. 

\begin{sidewaysfigure}[h]
	\subfigure[\label{Fig:OHPathDOS}]{\includegraphics[scale=0.5]{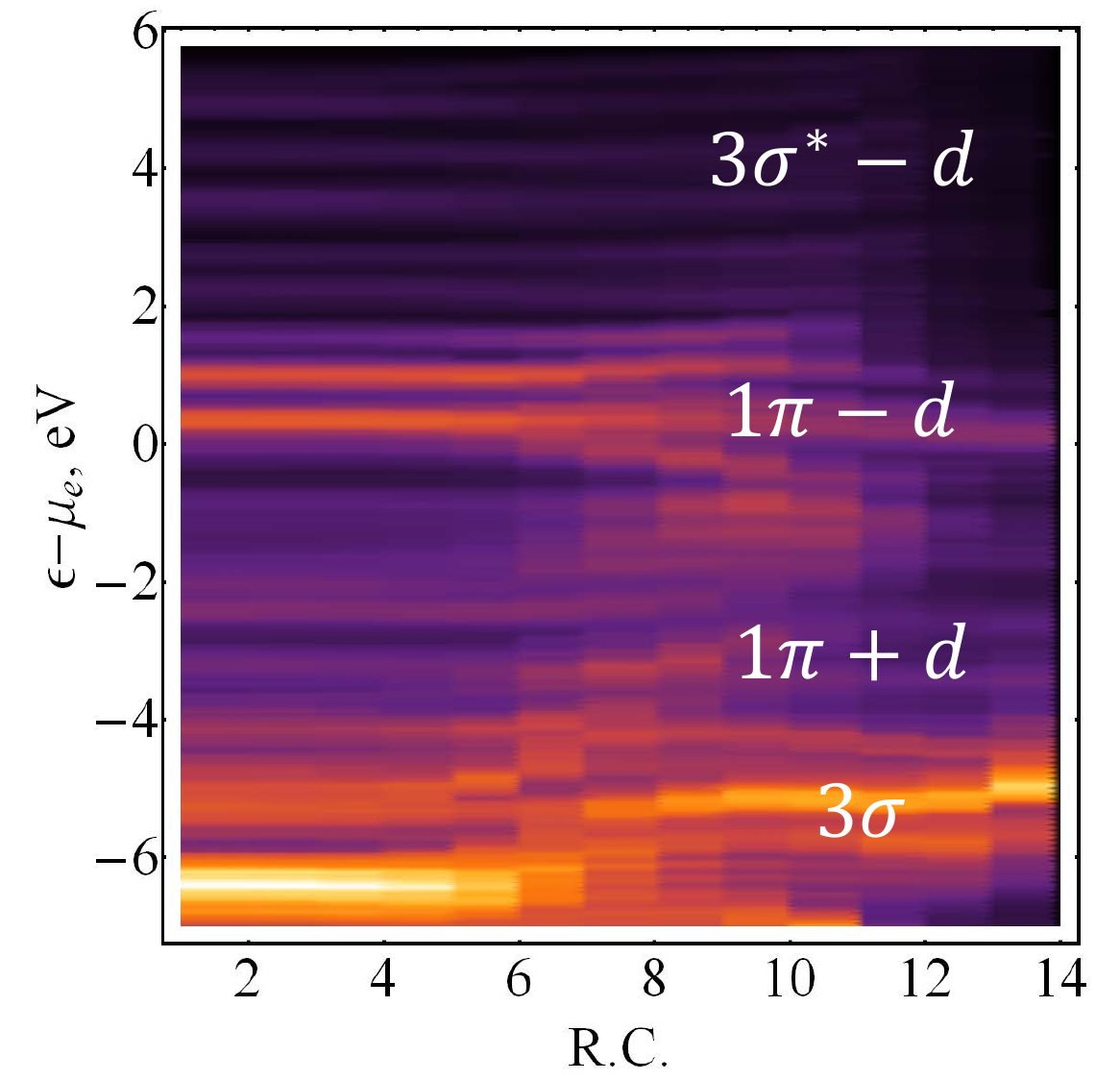}}
\qquad
\subfigure[\label{Fig:H2OPathDOS}]{\includegraphics[scale=0.5]{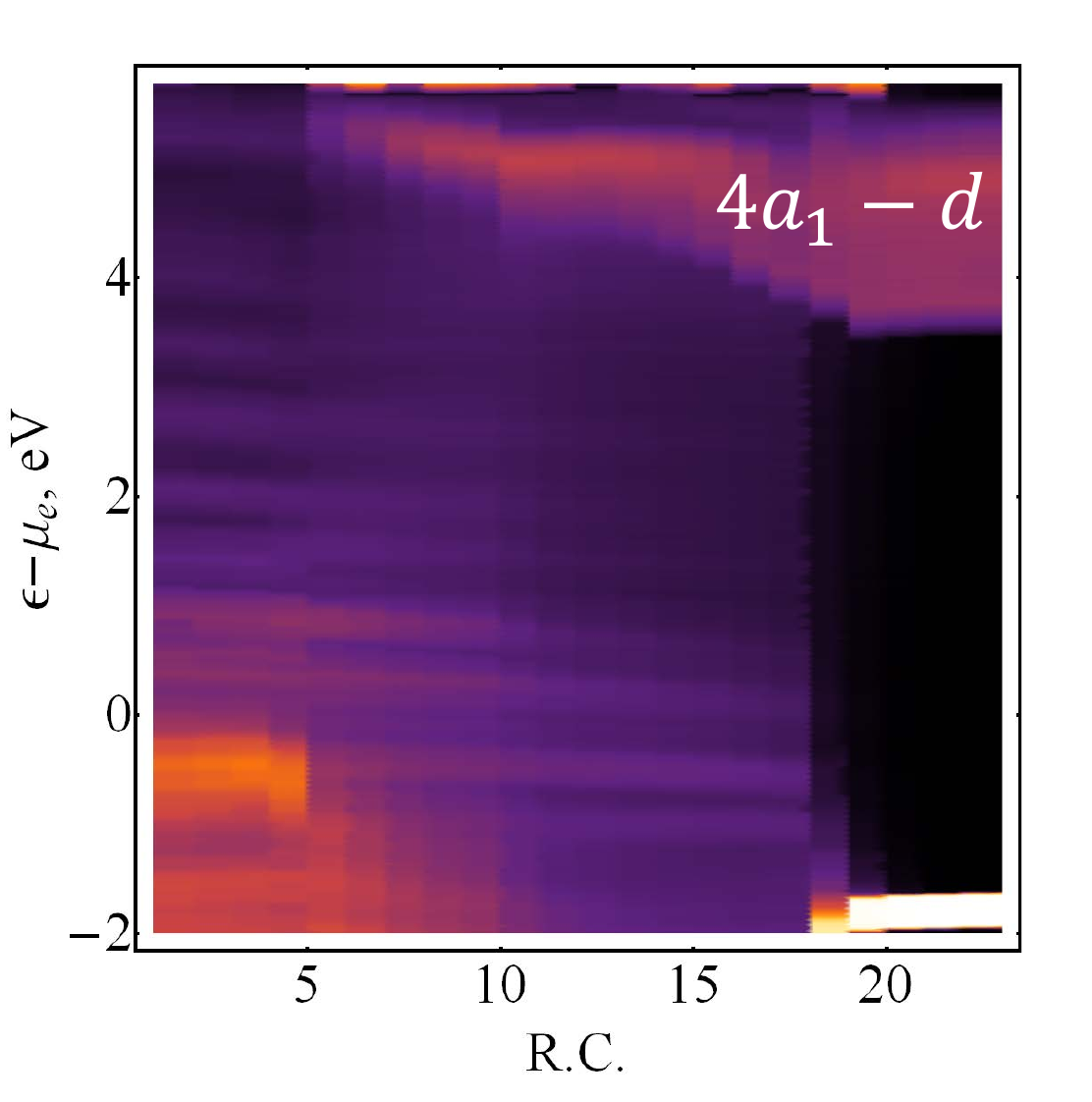}}
\caption
{
The density of the Kohn-Sham electronic states along (a) the segment of the reaction 
path describing  the event 
$\mathrm{``O^-+H^-"\rightarrow}``\mathrm{OH^-+e}"$, and 
(b) along the segment  $P_4$ of the reaction 
path describing the event 
$\mathrm{``OH^-+H^-"\rightarrow}``\mathrm{H_2O+2e}"$.
The vertical axis is the Kohn-Sham binding energy with respect to the Fermi level. 
The horizontal axis is the reaction coordinate. The brighter and darker colors correspond to higher and lower
values of the density of states, respectively. 
In (a), 
 $\sigma_{\sss{OH}}$ and $\sigma^*_{\sss{OH}}$ and $1\pi_{\sss{OH}}$ 
 hybridize with the platinum $\mathrm{5d6s}$ bands at the interface 
 to give rise to the split bonding and antibonding bands.
 In (b), the $4a_1-d$-derived unoccupied delocalized states can carry  the electrons that  previously resided within the occupied surface states. 
}
\end{sidewaysfigure}

As figs.~\ref{Fig:ABCD}(a,b) show,  the co-adsorbed $``\mathrm{O^-}"$ and $``\mathrm{H^-}"$ must overcome an activation barrier, 
$\chi^{\neq}_{\sss{OH(I)}}=0.79\;\mbox{eV}$,
during the up-hill step P$_2'$,
 before  $``\mathrm{OH^-}"$
 forms during the down-hill step  $P_2''$.
As \cref{Fig:OHPathDOS} indicates, the excess electron density  in \cref{Fig:ABCD}(a)
transforms from the initial $1s_{\sss{H}}\hyph$ and $2p_{\sss{O}}\hyph$like distribution to 
a bond-centered distribution during  $P''_{\sss{OH(I)}}$,
a fraction of the electrons redistributes to the non-bonding $1\pi_{\sss{OH}}$, where they remain, while
the rest return to the metal. The latter is indicated by the drop in the  charge at the adsorbates by $0.17\;\mbox{e}$ during
$P_2'$ and then by $0.24\;\mbox{e}$ during $P_2''$
(\cref{Fig:ABCD}(c)).
With the charge now concentrated at the non-bonding lone pairs  $1\pi_{\sss{OH}}$,
$``\mathrm{OH^-}"$ assumes the $\mathrm{O}$-end down  position (\cref{Fig:ABCD}(a)).
The energy that these electrons may take is estimated by the energy drop,
$\delta u_{\sss{OH(I)}}=1.1\;
\mbox{eV}$, during    $P_2''$. 

As \cref{Fig:ABCD}(a) shows,
$``\mathrm{H^-}"$ from an {\it fcc}
 hollow site migrates to a bridge site adjacent to the atop 
 $``\mathrm{OH^-}"$ during $P_4''$,
 and then moves towards the $``\mathrm{OH^-}"$ along the asymmetric stretch 
 of the nascent water molecule during $P_4'$.
 The energy in  \cref{Fig:ABCD}(b) increases towards the activation barrier, 
$\chi_{\sss{H_2O}}^{\ne}=0.3\;\mbox{eV}$,
 at $\mathrm{H_2O^{\ne}}$
  during $P_4'$.
 It drops by $
 \delta u^{\ne}_{\sss{H_2O}}$ (about $1.2\; \mbox{eV}$) 
to the minimum at $``\mathrm{H_2O}"$ during $P_4''$.
Meanwhile, the charge in  \cref{Fig:ABCD}(c) increases by
 $0.1\;\mbox{e}$
 during  $P_4'$, and then increases by $0.94\;\mbox{eV}$ during $P_4''$,
 indicating charge transfer to the metal. 
 This behavior is caused by
  the emptying of the continuum of
 the $\mathrm{H\cdots OH}$ LUMO ($4 a_1$)-derived
 states that host the excess charge before the transition configuration, 
$\mathrm{H_2 O^{\ne}}$ as  \cref{Fig:H2OPathDOS} illustrates. 
In other words,
 $\mathrm{H_2 O^{\sss{\ne}}}$
  is an 
 $\mathrm{``H\cdots OH^-}"$ that is just about to desorb while
 simultaneously autoionizing into the metal conduction band. 

Electron transfers in surface chemical events \cref{Eq:O2Diss0,Eq:H2Diss0,Eq:OHI0,Eq:H2O0} define  a surface electron pump, which  may contribute to 
the chemicurrent during catalytic  hydrogen oxidation \cref{hydrogen_oxydation_stoi}
on $\mathrm{Pt(111)}$ in a high temperature regime when water does not remain at the interface.
The legs  that form a cycle of the pump are 
  consequent reduction and oxidation  reaction. Delocalized conduction electrons from the metal localize at the interface along the reductive 
legs when surface anions form (i.e., in dissociative adsorption of hydrogen \cref{Eq:H2Diss0} and oxygen \cref{Eq:O2Diss0}). Then these localized electrons  return to the metal during the oxidative leg of the cycle when the surface anions autoionize (i.e., in the formation 
of  hydroxyl \cref{Eq:OHI0} and water \cref{Eq:H2O0}).
 It has long been established that metal electron can be excited in exothermic oxygen adsorption over  metal surfaces  \cite{doi:10.1142/S0218625X95000285}. Some of the heat released in this reaction is likely to excite metal electrons here.
Charge transfers from the interface  to the metal during the oxidative steps take place at transition states, which are high in  energy compared to
the  final states ($1.1-1.2\;\mbox{eV}$). Upon relaxation towards the final state, this excess energy must be shared between the rovibrational and electronic degrees of freedom, and, 
in the latter case, may channel to the chemicurrent in a Schottky diode with a low enough barrier. 
A possible energy transfer scenario involves  $4a_1$-derived states of a nascent water molecule, which carry   electrons localized at or before the anionic transition state
  to delocalized  metal conduction electrons above the Fermi level
at  the electroneutral product water molecule. 
This picture is reminiscent to that seen
in the context of catalytic $\mathrm{CO}$ oxidation on $\mathrm{Pt}(111)$\cite{ISI:000267972700008}, which also exhibits 
chemicurrent \cite{Somorjai:2005,Somorjai:2005:A} and involves charge transfer transition states.
Mechanism of \cref{hydrogen_oxydation_stoi} at low temperature when water is present at the surface is considerably more involved \cite{Fisher:Sexton:1980,Volkening:Jacobi:Ertl:1999,Michaelides:2001}. However, hydrated anions  must develop in this reaction, and these anions must change their charges during reaction steps (e.g.,$``\mathrm{(H_2O)_{n+1}O^-}+\mathrm{e}"\rightleftarrows``\mathrm{(H_2O)_{n}(OH)_2^{2-}}"$
and
$``\mathrm{(H_2O)_{n}(OH)_2^{2-}}+\mathrm{H^-}"\rightleftarrows``\mathrm{(H_2O)_{n+1}(OH)^{-}}+2\mathrm{e}"$
). These changes would  occur at transition states just as in CO oxidation and hydrogen oxidation at high temperature.
 
   This article  does not make any specific determination about the energy that these electrons carry;
a model beyond the adiabatic limit would be required to make such a determination and will be discussed elsewhere.
 However, the conclusion about the existence and importance of the charge transfer transition states in hydrogen oxidation, 
 as those configurations where the adiabatic picture fails in a particular way, 
 follows already from the structure of the ground state adiabatic potential.

\acknowledgements

The author thanks 
L.R. Backer, 
Y. Borodko,
 A. Hervier,
J.R. Renzas
G.A. Somorjai,
J.Y. Park
 for
comprehensive and stimulating discussions on the subject of their experimental work.
The author also thanks M. Head-Gordon for multiple stimulating discussions and providing critical comments on the manuscript.
This work was supported by the U.S. Department of Energy, Office of Basic Energy Sciences,
Chemical Sciences, Geosciences, and Biosciences Division, through Contract No. DEAC02-05CH11231.
This work
 used resources of
 the National Energy Research Scientific Computing Center.
 
 
\newpage

\newpage

\clearpage

\newpage

\clearpage

\newpage

\clearpage

\bibliography{H2O-Pt-fastelectrons-1}


%

\end{document}